\documentclass[12pt,letterpaper]{article}
 
\usepackage{amsmath}
\usepackage{amsfonts}
\usepackage{amssymb}

\usepackage{graphicx}

\def\be{\begin{equation}}
\def\ee{\end{equation}}
\def\ba{\begin{align}}
\def\ea{\end{align}}
\def\bear{\begin{eqnarray}}
\def\eear{\end{eqnarray}}
\def\nn{\nonumber}

\def\half{{{1\over 2}}}

\def\Heff{H_{\text{eff}}}

\begin{document}

%%%%%%%%%%%%%%%%%%%%%%%%%%%%%%%%%%%%%%%%%%%%%%%%%%%%%%%%%%%%%%%%%%%%
%  TITLE PAGE                                                      %
%%%%%%%%%%%%%%%%%%%%%%%%%%%%%%%%%%%%%%%%%%%%%%%%%%%%%%%%%%%%%%%%%%%%

\begin{titlepage}
\vskip 1in
\begin{center}
{\Large
{Noncommutative spaces and matrix embeddings on flat $\mathbb{R}^{2n+1}$}} % $\mathbb{R}^d$}}
\vskip 0.5in
{Joanna L. Karczmarek and Ken Huai-Che Yeh}
\vskip 0.3in
{\it 
Department of Physics and Astronomy\\
University of British Columbia\\
Vancouver, Canada}
\end{center}

\vskip 0.5in
\begin{abstract}
We conjecture an embedding operator which assigns, to any $2n+1$
hermitian matrices, a $2n$-dimensional hypersurface in flat $(2n+1)$-dimensional
Euclidean space.  This corresponds to precisely defining a fuzzy D(2n)-brane
corresponding to $N$ D0-branes.  Points on the emergent hypersurface correspond to 
zero eigenstates of the embedding operator, which have an interpretation
as coherent states underlying the emergent noncommutative geometry.
Using this correspondence, all physical properties of the emergent D(2n)-brane
can be computed.  We apply our conjecture to noncommutative flat and spherical
spaces.  As a by-product, we obtain a construction of a rotationally 
symmetric flat noncommutative space in 4 dimensions.
\end{abstract}
\end{titlepage}

%\tableofcontents

%%%%%%%%%%%%%%%%%%%%%%%%%%%%%%%%%%%%%%%%%%%%%%%%%%%%%%%%%%%%%%%%%%%%
%  BEGIN HERE                                                      %
%%%%%%%%%%%%%%%%%%%%%%%%%%%%%%%%%%%%%%%%%%%%%%%%%%%%%%%%%%%%%%%%%%%%

\section{Introduction and conjecture}
\label{intro}

The appearance of matrix coordinates, where the positions
of $N$ identical objects in $d$ dimensions are described by $d$
$N\times N$ matrices instead of $N$ $d$-vectors, is common in
string theory.  Geometric interpretation of non-commuting matrix coordinates
often involves an emergent higher dimensional object.  The exact shape
and other properties of this emergent object can be hard to study;
outside of highly symmetric surfaces such as spheres, only
some approximate methods (such as diagonalizing the matrices one at a times)
are usually employed.  In \cite{Berenstein:2012ts}, a method for determining a 
surface embedded in $\mathbb{R}^3$ and associated with any
three  matrices was given, providing a concrete
solution to this problem.  In \cite{deBadyn:2015sca}, the geometry of this surface
was examined in detail, proving the correspondence principle 
between matrix commutators and a Poisson structure on the emergent
surface.  

It is natural to ask about generalizing these results to higher
dimensions.  Higher dimensional noncommutative spaces posses a much
richer phenomenology than noncommutative surfaces do and
an explicit embedding into flat space would make their
study easier and more concrete.  Below, in equation (\ref{heff}), we 
conjecture an embedding operator which makes this possible for
even-dimensional noncommutative hypersurfaces embedded in a
odd-dimensional flat space.

In their paper, \cite{Berenstein:2012ts}  use a probe
brane interacting with a stack of $N$ D0-branes at an orbifold point
in the BFSS model, reducing the dimension of the space transverse to the
D0-branes to three. The emergent surface is defined as the locus of possible positions for the probe brane
where a fermionic string stretched from the stack to the probe brane 
has a massless mode. The fermion mass matrix is given by the following
effective Hamiltonian:
\be
\Heff(x_i) = \sum_{i=1,2,3} \sigma^i\otimes\left (X_i - x_i \right ) ~,
\label{BD}
\ee
where $\sigma^i$ are Pauli matrices, $X_i$ are Hermitian $N \times N$ matrices
corresponding to the positions of the stack of D0-branes and $x_i$ are
the positions of the probe brane. The stretched string has a zero mass fermionic mode 
when $\Heff$ has a zero eigenvalue.  Thus, the surface
corresponding to the three matrices $X_i$ is given by
the locus of points where $\Heff$ has zero eigenvalues,
defining a co-dimension one surface in flat $\mathbb{R}^3$.

$\Heff$ above plays a role of an `embedding operator':  it specifies
how the emergent surface given by three matrices $X_i$ should be embedded in
flat $\mathbb{R}^3$.
Since equation (\ref{BD}) was obtained from an orbifold construction, with
Pauli matrices arising from a dimensional reduction of Dirac $\Gamma$ matrices
from 9 dimensions to 3, a natural guess for the generalization of the
embedding operator to arbitrary odd dimensions is
\be
E_d(x_i) = \sum_{i=1}^d \gamma^i\otimes\left (X_i - x_i \right ) ~,
\label{heff}
\ee
where $\gamma^i$ are the (Euclidean) Dirac matrices in $d$ dimensions, which form
a representation of the Clifford algebra
\be
\{\gamma^i,\gamma^j\}=2\delta^{ij}~.
\ee
We have introduced a new symbol, $E_d$, to denote the embedding operator in
$\mathbb{R}^d$.  For $d=9$ this operator has been used in
\cite{Berenstein:2013tya} to study thermal configurations in the
BFSS model.  Similar Dirac operators have been used in 
\cite{Chatzistavrakidis:2011gs} to define the location of
D-brane intersections and the resulting emergent gravity.

As we will see, our conjectured  embedding operator 
`knows' a lot about noncommutative
geometry.  For example, a noncommutative sphere $S^{2d}$ with $SO(2d+1)$ symmetry
cannot locally (near some point $p$) look like the standard flat noncommutative space, 
since the latter is never fully rotationally symmetric, while $S^{2d}$
should retain $SO(2d)$ symmetry  around point $p$.  
Examining the kernel of the embedding operator $E_d$ for a noncommutative four-sphere
we find an auxiliary spin space whose presence restores $SO(4)$ invariance,
resolving the puzzle.  

It would be very interesting to obtain formula (\ref{heff})
from string theory considerations.  For $d=5$ and $d=7$, the computation 
might proceed along the lines of \cite{Berenstein:2012ts}, using an orbifold.  For $d=9$, another
method might be more applicable (see the discussion in \cite{Berenstein:2012ts}).  

The remainder of this paper is organized as follows: in the next section, 
we set conventions and observe that once  $E_d$ is known
in some odd dimension $d$, it is possible to obtain the embedding operators
in all lower dimensions by simply setting some of the matrices to zero, two at a time.
In section \ref{flat}, we discuss flat noncommutative space and generalize 
most of our results from \cite{deBadyn:2015sca} to higher dimensions.  In section \ref{spheres-even}
we study the noncommutative four-sphere embedded in $\mathbb{R}^5$, in 
particular obtaining a flat noncommutative space with $SO(4)$ rotational
symmetry as an approximation to the sphere on a small patch.
In section \ref{examples}, we discuss further examples of four dimensional
noncommutative surfaces.  Finally, in section
\ref{even} we try to study even dimensions by setting just one of the
matrices to zero. That this naive guess  fails to work can be demonstrated
by considering the noncommutative three-sphere, $S^3$.

\section{Conventions and a recursive property of $E_d$}
\label{notation}

Our embedding operators have the property that  once  $E_d$ is known
in some odd dimension $d$, it is possible to obtain the embedding operators
in all lower dimensions recursively. 
To easiest way to see that our family of embedding operators has this property is to use an
iterative definition of the $\gamma$ matrices as follows.\footnote{We follow here \cite{Polchinski:1998rr}.}
In $d=1$, we trivially take $\gamma^1$ to be the $1\times1$ unit matrix.
Then, denoting the
$\gamma$ matrices in $d-2$ dimensions with $\tilde\gamma^i$, we have in $d$
dimensions that
\begin{align}
\gamma^i&=\sigma^3\otimes\tilde\gamma^i~,~ ~i=1,\ldots,d-2~, \\
\gamma^{d-1} &=   \sigma^1\otimes\boldsymbol{1}~, \\
\gamma^{d} &=  \sigma^2\otimes\boldsymbol{1}~.
\end{align}
The dimension of the $\gamma$ matrices is thus $2^n=2^{(d-1)/2}$.
For $d=3$, we obtain a permutation of the Pauli matrices:
$\gamma^1 = \sigma^3$, $\gamma^2 = \sigma^1$, $\gamma^3 = \sigma^2$.

Now,  in some odd number of dimensions $d$ set  the last two matrices $X_{d-1}$ and
$X_{d}$ to zero.  We can then reduce $E_{d}$ to
$E_{d-2}$: if $X_{d-1}=X_{d}=\boldsymbol{0}$, then
\ba
E_{d}(x_1,\ldots,x_{d}) 
&=\sum_{i=1}^{d-2} ~ \left(\sigma^3\otimes\tilde\gamma^i\right)
\otimes \left (X_i - x_i \right ) 
- (\sigma_1\otimes\boldsymbol{1})\otimes(x_{d-1})%\boldsymbol{1})
- (\sigma_2\otimes\boldsymbol{1})\otimes(x_{d})%\boldsymbol{1})
%\nn\\
%&=\sigma^3\otimes E_{d-2}(x_1,\ldots,x_{d-2}) ~.
\end{align}
One can show that for the above operator have a zero eigenvector,
we must necessarily have $x_{d-1}=x_d=0$.  Then, the operator above
can be reduced to
\be
\sigma^3 \otimes E_{d-2}(x_1,\ldots,x_{d-2})~. 
\ee
Thus, once a construction of $E_d$ is known in some odd number of
dimensions, it is easy to construct all the smaller odd dimensional cases.
In section \ref{even} we will discuss our attempt to obtain an embedding operator
in an even number of dimensions by setting just one of the matrices to zero.

To write the $\gamma$ matrices in an explicit form it is convenient
to introduce the following notation: 
\be
\sigma_{n}(c_1,\ldots,c_n):= \sigma^{c_1} \otimes ~\ldots~\otimes \sigma^{c_n}~,
\ee
where the coefficients $c_i$ take integer values from 0 to 3 and where we
define $\sigma^0 = \boldsymbol{1}$.  
%We will also use $c_i = \pm$ with the obvious intepretation.
In this notation, the recursive definition of $\gamma$ matrices implies that
\ba
\gamma^{1} &= \sigma_{n}(3,3,3,\ldots,3,3,3) \nn\\
\gamma^{2} &= \sigma_{n}(3,3,3,\ldots,3,3,1) \nn\\
\gamma^{3} &= \sigma_{n}(3,3,3,\ldots,3,3,2) \nn\\
\gamma^{4} &= \sigma_{n}(3,3,3,\ldots,3,1,0) \nn\\
\gamma^{5} &= \sigma_{n}(3,3,3,\ldots,3,2,0) \nn\\
%\gamma^{6} &= \sigma_{n}(3,3,3,\ldots,1,0,0) \nn\\
%\gamma^{7} &= \sigma_{n}(3,3,3,\ldots,2,0,0) \nn\\
&\vdots\nn\\
%\gamma^{d-3} &= \sigma_{n}(3,1,0,\ldots,0,0,0) \nn\\
%\gamma^{d-2} &= \sigma_{n}(3,2,0,\ldots,0,0,0) \nn\\
\gamma^{d-1} &= \sigma_{n}(1,0,0,\ldots,0,0,0) \nn\\
\gamma^{d} &= \sigma_{n}(2,0,0,\ldots,0,0,0) \nn
\end{align}

To complete our conventions, we make the following choice for the Pauli matrices:
\be
\sigma^1 = \begin{bmatrix}
0 & 1 \\ 1 &0
\end{bmatrix}~,~~~
\sigma^2 = \begin{bmatrix}
0 & -i \\ i &0
\end{bmatrix}~,~~~
\sigma^3 = \begin{bmatrix}
  1 & 0 \\ 0 &-1
\end{bmatrix}~,~~~
\sigma^- = \begin{bmatrix}
0 & 0 \\ 1 &0
\end{bmatrix}~,~~~
\sigma^+ = \begin{bmatrix}
0 & 1 \\ 0 &0
\end{bmatrix}
~.
\ee

\section{Noncommutative ${\mathbb{R}}^{2n}$}
\label{flat}

As the first example, set $X^1=\boldsymbol{0}$ and consider 
the other $d-1$ matrices to have a commutation relation
\be
[X_i,X_j] = i\theta_{ij}~\text{for}~i,j=2,\ldots,d~.
\label{flat-nc-general}
\ee
This, of course, is simply flat noncommutative space, extending in
dimensions $2$ through $d$ (assuming $\theta$ has full rank).  $\theta$ is an antisymmetric 
even dimensional matrix which can be, by an orthogonal change of basis and 
therefore without loss of generality, brought
into the block-diagonal form 
\be\theta =\text{diag}\left (
\left [ \begin{array}{cc}
0 & \theta_1  \\ -\theta_1  & 0
\end{array} \right ],~\ldots,~
\left [ \begin{array}{cc}
0 & \theta_n  \\ -\theta_n  & 0
\end{array} \right ] \right )~.
\label{flat-nc-diagonal}
\ee
We define $A_a = X_{2a} + i X_{2a+1}$ for $a=1,\ldots,n$.
$A_a$ and $A_a^\dagger$ are the lowering and raising operators
of a harmonic oscillator with $[A_a,  A_a^\dagger] = 2\theta_a$.
The lowering operators $A_a$ have
eigenstates $|\alpha \rangle_a$ (the coherent states),
corresponding to  every complex number $\alpha$: 
$A_a |\alpha \rangle_a =  \alpha |\alpha \rangle_a$.  
$E_d$ can be written as 
\be
\sum_{a=1}^n \left( \Lambda^a_-\otimes(A_a-\alpha_a) +
\Lambda^a_+\otimes(A_a^\dagger-\bar\alpha_a)\right)~,
\ee
where
\be
\Lambda^a_\pm ~=~ %\left(\sigma^3\right)^{\otimes(n-a)}~\otimes~\sigma_\pm~
%\otimes \left (\sigma^0\right)^{\otimes(a-1)}~ =~ 
\sigma_n(\underbrace{3,\ldots,3}_{n-a\text{~times}},~\pm,~
\underbrace{0,\ldots,0}_{a-1\text{~times}})~=~\gamma_{2a} \pm i \gamma_{2a+i}~.
\label{raising-lowering-gamma}
\ee
In this form, it is easy to see that
\be
|\Lambda(\alpha) \rangle  =  \left (\bigotimes_{a=1}^n 
\left [ \begin{array}{c} 1 \\  0 \end{array}\right ] \right )~\otimes~
\left ( \bigotimes_{a=1}^n|\alpha\rangle_a \right )
\ee
is a zero eigenvector for $E_d$ at a point given by $x_1=0$ and $x_{2a}+ix_{2a+1}=\alpha_a$.
Thus there is a zero eigenvector for every point on the co-dimension 
one hypersurface given by $x_1=0$.  The first factor in
the above zero eigenvector is simply one of the highest weight vectors
of the Clifford algebra 
selected by the particular form of raising operators $\Lambda_+$ which we are using.  
We will denote it with $V_d$:
\be
V_d := \left (\bigotimes_{a=1}^n 
\left [ \begin{array}{c} 1 \\  0 \end{array}\right ] \right )~.
\ee
$V_d$ is an eigenvector of $\gamma^1$ with eigenvalue 1 and 
has the property that $\Lambda_+^a V_d = 0$ for all  $a$.

We can expect many noncommutative spaces to have the property
that the embedding operator has a single zero eigenvector at a given
point on the emergent surface.  Those spaces should, 
locally, look like noncommutative flat space given by equation (\ref{flat-nc-general}).    
We will see in the next section that, for $d>3$, non-degenerate noncommutative
spaces exist whose embedding operators have multiple zero eigenvectors at a point.
However, for those that don't, our work \cite{deBadyn:2015sca} on emergent surfaces 
in the large $N$ limit
can easily be generalized to higher dimensions.  
Similar results have been obtained before in \cite{Steinacker:2010rh}
(see also \cite{Steinacker:2011ix} and the references therein).

In the rest of this section,
we state the salient results and conjectures.

Assume, then, that the zero eigenvector $|\Lambda_p\rangle$ of the embedding operator is unique at every
point $p$ of the emergent surface.  The normal vector to this surface at point $p$ is
given by\footnote{The arguments for this and other statements below are
basically identical to that given in \cite{deBadyn:2015sca} for $d=3$.}
\be
n_i = \langle \Lambda_p | \gamma_i \otimes \boldsymbol{1} | \Lambda_p\rangle~.
\ee
For simplicity, we now rotate our surface so that the normal vector at the 
point of interest points in the $x_1$ direction.
We conjecture that the eigenvector is equal to, approximately,  a product of an 
appropriate highest weight state  $V_d^\theta$ and a $N$-dimensional
vector:
\be
|\Lambda\rangle ~=~ V_d^\theta \otimes |\alpha\rangle ~+~\text{corrections that vanish for }N\rightarrow 0~.
\label{factorization}
\ee
Further, we can define a local noncommutativity matrix $\theta_{ij}$ at point $p$ by 
\be
\theta_{ij}=\langle \alpha | -i[X_i,X_j] | \alpha \rangle~,~~\text{for }i,j = 2,\ldots d.
\ee
$\theta_{ij}$ is an antisymmetric two-form on the emergent surface; it defines
 a Poisson bracket of two functions $f$ and $h$:
\be
\{f,h\} ~:=~ \frac{N\theta_{ab}}{\sqrt{\det g}}~ \partial_a f \partial_b h~,
\label{poisson}
\ee
where $g_{ab}$ is the pullback of the flat metric on $\mathbb{R}^{d}$ to the $d-1$ dimensional
emergent space.  From this Poisson bracket, we divide the $d-1$ directions $x_2,\ldots,x_d$
into raising and lowering operators just like we did above.  In particular, we have
a new set of lowering and raising operators on the spinor space, $\Lambda^{\theta,a}_\pm$
(defined, in a particular basis, in equation (\ref{raising-lowering-gamma})).  
The highest weight state in equation (\ref{factorization}) has
$\Lambda^{\theta,a}_+ V_d^\theta = 0$.

The $N$-dimensional state $|\alpha\rangle$ should be interpreted as a coherent
state associated with the point $p$.  Since $(E_d)^2|\Lambda\rangle = 0$, we have
\be
\langle \Lambda_p|\boldsymbol{1} \otimes   \sum_i (X_i- x_i)^2 
| \Lambda_p \rangle = 
- \half   \langle \Lambda_p | \sum_{i\neq j}\left ( \gamma_i\gamma_j)\otimes[X_j,X_k] \right )| \Lambda_p \rangle~. 
\ee
Substituting the factorization condition (\ref{factorization}), we obtain
\be
\langle \alpha | \sum_i (X_i- x_i)^2  |\alpha \rangle = 
- \half   \langle \alpha | \sum_{i\neq j}\left (n_{ij}[X_i,X_j] \right )| \alpha \rangle~, 
\label{coherent}
\ee
where the two-form $n_{ij}$ is defined below, in equation (\ref{2-form-n}).
However, since our noncommutative space is a direct product of $n$ copies
of two dimensional noncommutative space, a better way to study the properties
of the coherent state is work in a basis where the noncommutativity is given
by equation (\ref{flat-nc-diagonal}) and to write $|\alpha\rangle$ as a product of $n$ coherent states
$|\alpha\rangle = |\alpha\rangle_1 \otimes \ldots \otimes |\alpha\rangle_n$.

Once we have coherent states $|\alpha_p\rangle$ corresponding to every point $p$ on the surface,
we can associate any $N\times N$ matrix $M$ with functions on the surface, via 
$M \rightarrow \langle\alpha_p |M |\alpha_p\rangle$.  This is the Berezin approach to
noncommutative geometry \cite{Berezin:1974du}.  It gives a natural map between commutators
of operators and an antisymmetric Lie bracket on the surface.  This bracket turns
out to be equal to the Poisson bracket defined in equation (\ref{poisson}) as long as,
in addition to the factorization condition (\ref{factorization}), we also have that
\be
\langle \alpha | -i[X_j,X_1] | \alpha \rangle~,~~\text{for }i,j = 2,\ldots d,
\ee
is much smaller than $\|\theta_{ij}\|$ for $N\rightarrow \infty$.  

It is useful to define two antisymmetric two-forms on $\mathbb{R}^d$:
\be
\hat \theta_{ij}=\langle \alpha | -i[X_i,X_j] | \alpha \rangle~,~~\text{for }i,j = 1,\ldots d.
\label{2-form-theta}
\ee
and
\be 
n_{ij} = \half \langle V_d^\theta | i [\gamma_i,\gamma_j] | V_d^\theta \rangle \text{ for } i \neq j~.
\label{2-form-n}
\ee
It is easy to see that $n_{1k} = 0$.
In the basis in which $\theta_{ij}$ is given by equation (\ref{flat-nc-diagonal}), we have
\be n_{ij} =\text{diag}\left ( 0, 
\left [ \begin{array}{cc}
0 & 1  \\ -1  & 0
\end{array} \right ],~\ldots,~
\left [ \begin{array}{cc}
0 & 1  \\ -1  & 0
\end{array} \right ] \right )~.
\ee
The necessary condition for the correspondence principle to hold can then be stated more covariantly as
\be
n_{ij} n_{kl} \hat\theta_{ik} = \theta_{jl} ~+~\text{corrections that vanish for }N\rightarrow 0~.
\ee
It follows that the vector $\epsilon^{i_1,i_2,\dots,i_d}\hat\theta_{i_1,i_2} \ldots\hat\theta_{i_{d-2},i_{d-1}}$
should be nearly parallel to the normal vector $n_i$.  We conjecture that this vector is 
related to the total volume of the surface via
\be
\text{Volume}_{d-1}(\text{emergent surface}) = C~\text{Tr} ~\sqrt{
\sum_{i_d} \left (\epsilon^{i_1,i_2,\dots,i_d} [X_{i_1},X_{i_2}]\ldots [X_{i_{d-2}},X_{i_{d-1}}] \right )^2}~.
\label{volume}
\ee
$C$ in the above is some numerical coefficient which does not depend on $N$ (for
$d=3$, this coefficient was $2\pi$).

When interpreting the emergent  surface as a higher-dimensional D-brane
emerging from D0-branes via the dielectric effect \cite{Myers:1999ps},
the two form $\hat \theta_{ij}$ and its pullback to the worldvolume of the
D-brane, $\theta_{ij}$, will enter into the non-abelian BI and CS actions
as expected.  Finally, an emergent D-brane should have a $U(1)$ connection
living on its worldvolume; following \cite{Berenstein:2012ts}, we can define it as
\be
2 v^iA_i = -iv^i\langle \alpha(x_i) |\partial_i|\alpha(x_i)\rangle ~,
\label{connection-1}
\ee
where $v^i$ is a tangent vector on the emergent surface.  Working with a
coherent state in a factorized form, we obtain that associated curvature is $\partial_{[i}A_{j]}
=(\theta^{-1})_{ij}$, as expected.

\section{Even dimensional spheres $S^{2n}$ and noncommutative space with $SO(2n)$ invariance}
\label{spheres-even}

The noncommutative four sphere can be constructed as in \cite{Grosse:1996mz}
(see also \cite{Castelino:1997rv}).   The starting point is a representation
of the Clifford algebra in four dimensions: the $\gamma$ matrices
of section \ref{intro}.  The matrices in this representation act on vectors
in a four-dimensional spinor representation.  Consider then an
irreducible representation of $Spin(5)$ given by the completely symmetric
tensor product of $k$ copies of this irrep.  To each $\gamma$, associate
a matrix $X^i$ that acts on this tensor product as follows
\be
X_i ~=~ \frac{1}{k} \, (\gamma^i\otimes\boldsymbol{1}\otimes\ldots\otimes\boldsymbol{1}~+~
\boldsymbol{1}\otimes\gamma^i\otimes\ldots\otimes\boldsymbol{1}~+~\ldots~+~
\boldsymbol{1}\otimes\boldsymbol{1}\otimes\ldots\otimes\gamma^i)_\text{sym}
\label{sphere-def}
\ee
The claim is that these five position matrices represent a four-sphere of radius one.
Their dimension is $N=(k+1)(k+2)(k+3)/6$.

In the four dimensional spinor irrep, consider the the vector $V_4$, and
take its image under the $k^\text{th}$ symmetric tensor product map, 
$(V_4\otimes\ldots\otimes V_4)_\text{sym}:=(V_4)^{\otimes k}$.
Because $\gamma^1 V_4 = V_4$ and $(\gamma^2+i\gamma^3)V_4 = (\gamma^4+i\gamma^5)V_4 = 0$,
the matrices $X_i$ above have a simple action on this vector,
\ba
X_1\,\cdot\, (V_4)^{\otimes k} &= (V_4)^{\otimes k}~, \\
(X_2 + i X_3) \,\cdot\,(V_4)^{\otimes k} &= 0~,\\
(X_4 + i X_5)\,\cdot\, (V_4)^{\otimes k} &= 0~.
\end{align}

Consider the point $(1,0,0,0,0)$ in $\mathbb{R}^5$, which we hope lies
on the emergent sphere.  We need the embedding operator $E_5$ at this point to have a zero eigenvector.
We rewrite $E_5$ at this point as
\be
\gamma^1\otimes(X^1-1) ~+~ \sum_{a=1}^2 \left( \Lambda^a_-\otimes X_a^+ +
\Lambda^a_+\otimes X_a^-\right)~,
\label{heff-5-sphere}
\ee
where
\be
X_a^{\pm} = X_{2a} \pm iX_{2a+1}~.
\ee
Now, consider a vector $\Lambda = V_4 \otimes (V_4)^{\otimes k}$.  It is easy
to see that this is a zero eigenvector of $E_5$ as given in equation (\ref{heff-5-sphere}).
Since the $\gamma$ matrices form a fundamental (or standard) representation of
$so(5)$, we recover the entire spherical surface with radius 1 by symmetry.  However,
using the methodology from section \ref{flat}, rotational symmetry appears lost, as 
$\langle (V_4)^{\otimes k} | [X_2,X_3] |(V_4)^{\otimes k}\rangle = 
\langle (V_4)^{\otimes k} | [X_4,X_5] |(V_4)^{\otimes k}\rangle = 1/k$
and the other four commutators vanish.  Since we know that the noncommutative sphere
has SO(5) symmetry and therefore SO(4) symmetry once a point on the sphere is fixed,
the noncommutativity on the 4-sphere must be of a different kind than that in section
\ref{flat}.

In fact,
%An attentive reader might have noticed that 
$\Lambda = V_4 \otimes (V_4)^{\otimes k}$
is not the only zero eigenvector of the embedding operator in equation (\ref{heff-5-sphere}).
In contrast to the flat noncommutative space above, here both the raising and the
lower operators $X_a^\pm$ have zero eigenvectors.  Let $\tilde V_4 = \Lambda_-^1 \Lambda_-^2 V_4$ 
be the spinor with $\gamma^1 \tilde V_4 = \tilde V_4$  and $\Lambda_-^a \tilde V_4 = 0$.
Then, consider an arbitrary unit spinor $W$ in the span of $\{V_4,\tilde V_4\}$,
$W = \mu V_4 - \nu \tilde V_4$, $|\mu|^2+|\nu|^2=1$, with $\gamma^1 W = W$,
$(\mu\Lambda_+^1 - \nu \Lambda_-^2)W=0$ and  $(\mu\Lambda_+^2 + \nu \Lambda_-^1)W=0$.
Rewrite the embedding operator in equation (\ref{heff-5-sphere}) as
\ba
& \gamma^1\otimes(X^1-1) ~+~  \\
&\left( \mu\Lambda^1_+ -\nu\Lambda_-^2 \right) \otimes  \left(\mu X_1^- - \nu X_2^+\right) ~+~ 
\left( \nu\Lambda^1_+ + \mu\Lambda_-^2 \right) \otimes  \left(\nu X_1^- + \mu X_2^+\right) ~+~ \nn \\ \nn
%&\text{      exchange }1\text{ and } 2\text{ on the previous line}
&\left( \mu\Lambda^2_+ -\nu\Lambda_-^1 \right) \otimes  \left(\mu X_2^- - \nu X_1^+\right) ~+~ 
\left( \nu\Lambda^2_+ + \mu\Lambda_-^1 \right) \otimes  \left(\nu X_2^- + \mu X_1^+\right) ~.
\label{heff-5-sphere-arbitrary}
\end{align}
This demonstrates explicitly that
\be
\tilde \Lambda = W \otimes  (W)^{\otimes k}~
\ee
is also a zero eigenvector of the embedding operator in equation (\ref{heff-5-sphere}).  
The kernel of the embedding operator is a $(k+2)$-dimensional space, while
the space of the associated coherent states is $(k+1)$-dimensional.\footnote{
First, let's understand why the vectors $(W)^{\otimes m}$ span a $(m+1)$-dimensional space
(ie, why $\text{Sym}^m(\text{span}\{V_4,\tilde V_4\})$ is $(m+1)$-dimensional),
by drawing a parallel with representations of $SU(2)$.  The fundamental
irrep of $SU(2)$ is of course 2-dimensional, and all higher irreps
correspond to completely symmetric tensor powers of the fundamental representation.
Thus we know that the dimension of $\text{Sym}^m S$ where $S$ is any two
dimensional vector space is $m+1$.  Thus, the kernel has dimension $k+2$ (because
it corresponds to $W^{\otimes(k+1)}$), but there are only $k+1$ linearly
independent $N$-dimensional coherent states once the first term in the product 
is stripped off.}
Its presence has a natural interpretation:
it is the auxiliary space necessary to ensure that the emergent
noncommutative space has $SO(4)$ symmetry.  (Notice that the noncommutative
flat space we defined in the previous section {\it{does not}} have full rotational
symmetry even when we set all $\theta_{a}$ equal to each other.)  

To see how rotational invariance is restored, first notice that the SO(4) symmetry
we wish to see restored is generated by the six commutators $[X_i,X_j]$.  For $k=1$,
we write these commutators explicitly:
\ba
&L_1:=-i[X_2,X_3] = \sigma_2(0,3)~,~&L_2:=-i[X_2,X_4] = \sigma_2(2,1)~,~&L_3:=-i[X_3,X_4] = \sigma_2(2,2)~,~\nn\\
&K_1:=-i[X_4,X_5] = \sigma_2(3,0)~,~&K_2:=-i[X_5,X_3] = \sigma_2(1,2)~,~&K_3:=-i[X_2,X_5] = -\sigma_2(1,1)~.\nn
\end{align}
For larger $k$, we just consider these operators acting on the symmetric k$^\text{th}$ tensor
power of the four-dimensional irrep of $Spin(5)$.
%, just like we did to get $X_i$ out of $\gamma_i$ in equation (\ref{sphere-def}).  
Notice than when one of these six generators acts on
any vector in the kernel of the embedding operator, we get another vector in the kernel.  
Thus, we get a representation of of the algebra ${so}(4)$.  To see which representation it is, 
consider two mutually commuting sets of generators, $L_i\pm K_i$.  Their commutation
relationships are
\be
[(L_i\pm K_i),(L_j\pm K_j)] = 2i\epsilon_{ijk} (L_k\pm K_k) \text{ and }
[(L_i\pm K_i),(L_j\mp K_j)] = 0~,
\ee
which is nothing more but the standard fact that $SO(4) \sim SU(2) \times SU(2)$.
By explicit computation, we see that when acting on the kernel of the embedding operator,
$L_i-K_i$ vanish, while the action of $L_i+K_i$ is that of a (k+1)-dimensional irreducible 
representation of ${su}(2)$.  Thus, the zero
eigenvectors of the embedding operator form the $(k/2,0)$  irrep of $SU(2) \times SU(2)$.
For example, for $k=1$ we have, explicitly in the $\{V_4$, $\tilde V_4\}$ basis
\ba
-i[X_2,X_3] = \sigma_2(0,3) \rightarrow m_{23}:=\left [ \begin{array}{cc} 1 &  0 \\ 
0 & -1 \end{array}\right ]~, \nn \\
-i[X_3,X_4] = \sigma_2(2,2) \rightarrow m_{34}:=\left [ \begin{array}{cc} 0 &  -1 \\
-1 & 0 \end{array}\right ]~, \label{matrices}\\
-i[X_2,X_4] = \sigma_2(2,1) \rightarrow m_{24}:=\left [ \begin{array}{cc} 0 &  i \\ 
-i & 0 \end{array}\right ]~. \nn 
\end{align}

So far, we have focused on the point $(1,0,0,0,0)$.  However, when other points
close enough to this one are considered, the commutators $[X_i,X_j]$ for $i,j=2,\ldots,5$
are nearly constant.  Consider, for example, a zero eigenvector of the embedding operator
$E_5$ at a point $(1,\hat \beta,0,0,0)$, with $\beta \ll 1$.  
Let's use  a basis for the four dimensional spinor representation
given by $\sigma_2(3,0)|s_1,s_2\rangle = s_1 |s_1,s_2\rangle$ and 
$\sigma_2(0,3)|s_1,s_2\rangle = s_2 |s_1,s_2\rangle$, where $s_i = \pm 1$.  In this notation,
$V_4 = |\!+\!+\!\rangle$ and $\tilde V_4 = |\!-\!-\!\rangle$. For clarity, pick an eigenvector of the embedding
operator at point $(1,\hat \beta,0,0,0)$ of the form
\bear
&&(|\!+\!+\!\rangle ~+~ \beta |\!+\!-\!\rangle + \ldots)^{\otimes k} ~=~ \\ \nn 
&&|\!+\!+\!\rangle^{\otimes k} ~+~ k\beta (|\!+\!-\!\rangle\otimes|\!+\!+\!\rangle^{\otimes (k-1)})_\text{sym}
\\ \nn 
&&+ \half k(k-1)\beta^2 (|\!+\!-\!\rangle\otimes|\!+\!-\!\rangle\otimes|\!+\!+\!\rangle^{\otimes (k-2)})_\text{sym}
~+~ \ldots
\eear
where $\beta$ is proportional to $\hat \beta$.  
$(|\!+\!-\!\rangle\otimes|\!+\!+\!\rangle^{\otimes (k-1)})_\text{sym}$ has length
$1/\sqrt{k}$, $(|\!+\!-\!\rangle\otimes|\!+\!-\!\rangle\otimes|\!+\!+\!\rangle^{\otimes (k-2)})_\text{sym}$
has length approximately $1/\sqrt{k(k-1)/2}$, etc...,
Thus, $\sqrt{k}\beta$ is of order 1, this vector's overlap with $(|\!+\!+\!\rangle)^{\otimes k}$ decreases
sharply to zero\footnote{From our work \cite{deBadyn:2015sca}, we would expect this overlap to have Gaussian fall-off.}\footnote{
Thus, the radius of a noncommutative `cell' is $1/\sqrt{k}$ and its 4-volume is $1/k^2$.  In a sphere
of radius 1, we then have approximately $k^2$ such `cells'.  Each corresponds to
a $k+2$ dimensional kernel of the embedding operator, so the total dimensionality 
of the matrices needs to be approximately $k^3$, in agreement with the exact
formula $N=(k+1)(k+2)(k+3)/6$}.
That a smooth sphere is recovered in the large $k$ limit tell us that 
there is a range of values for $\beta$ (or $\hat \beta$) where the vector above is close to being
linearly independent of $(|\!+\!+\!\rangle)^{\otimes k}$ but where terms with powers of $\beta$
greater than some $p \ll k$ can be ignored.  In this range, the matrix elements of $[X_i,X_j]$
when acting on the kernel of the embedding operator are approximately independent of $\beta$.
As an example, consider that
\be
\langle\!+\!+\!|^{\otimes k}~ -i[X_3,X_4] ~~(|\!+\!-\!\rangle\otimes|\!+\!+\!\rangle^{\otimes (k-1)})_\text{sym}
\ee
is of order $1/k$, because the above overlap is only nonzero when the nontrivial
operator in 
\be
-i[X_3,X_4] = 
\frac{1}{k} \, (\sigma_2(2,2)\otimes\boldsymbol{1}\otimes\ldots\otimes\boldsymbol{1}~+~
\boldsymbol{1}\otimes\sigma_2(2,2)\otimes\ldots\otimes\boldsymbol{1}~+~\ldots~+~
\boldsymbol{1}\otimes\boldsymbol{1}\otimes\ldots\otimes\sigma_2(2,2))_\text{sym}
\ee
`finds' $|\!+\!\!-\!\rangle$ when acting on 
$(|\!+\!\!-\!\rangle\otimes|\!+\!\!+\!\rangle^{\otimes (k-1)})_\text{sym}$.

Thus, for points near $(1,0,0,0,0)$, the relevant commutators, when acting on the kernel of
the embedding operator, are nearly constant (with $1/k$
corrections) and we get the following approximate noncommutative algebra
\be
[X_i,X_j] = \frac{i}{k} m_{ij} ~,
\label{spin-nc}
\ee
where $m_{ij}$ are $(k+1)\times(k+1)$ matrices in the $(k/2,0)$ irreducible representation 
of $SU(2) \times SU(2)$.  $m_{23}$, $m_{24}$ and $m_{34}$ are defined in equation
(\ref{matrices}), while $m_{25}=m_{34}$, $m_{35}=m_{24}$ and $m_{45}=m_{23}$.
The factor $1/k$ comes from normalization of $X_i$ in equation (\ref{sphere-def}).  
This  noncommutativity algebra
is similar to spin noncommutativity with SO(3) symmetry in three spacial
dimensions in \cite{Falomir:2009cq} (see also references therein).

$SO(4)$ is restored in equation (\ref{spin-nc})  because the action of $SO(4)$ on 
$X_2, X_3, X_4, X_5$ can be `undone' by a similarity transformation
on matrices $m_{ij}$.  Since $SU(2)\times SU(2)$ is a double cover
of $SO(4)$, a rotation in ${\mathbb{R}}^4$ that goes `all the way around'
(ie, is trivial in $SO(4)$) corresponds to a nontrivial element of 
$SU(2)\times SU(2)$, namely $(-\boldsymbol{1})\otimes(-\boldsymbol{1})$.
In the $(1/2,0)$ irrep (and all $(k/2,0)$ irreps for $k$ odd) this corresponds
to multiplying all the vectors in the kernel of the embedding operator by $-1$.  
Such a change of basis has no effect on the matrix elements of $[X_i,X_j]$, or on 
$m_{ij}$. For $(k/2,0)$ irreps with $k$ even, $(-\boldsymbol{1})\otimes(-\boldsymbol{1})$
is trivial.

Another observation concerns orientability: a noncommutative 4-space
with opposite orientation to the one we have considered is found at the other
pole of the sphere, near the point $(-1,0,0,0,0)$.  This can be obtained by
taking $V_4 \rightarrow \tilde V_4$ and $\tilde V_4 \rightarrow  V_4$.
As such a map is not an element of $SU(2)$, it has a nontrivial effect
on the matrices $m_{ij}$

That the space of coherent states  has dimension $k+1$ fits well
with string theory: in \cite{Castelino:1997rv} it was found that the correct
interpretation of the four-sphere is that of a $D4$-brane stack with $k$
overlapping branes\footnote{Up to corrections of order $1/k$, which
explains the discrepancy between $k$ and $k+1$.}.  Further, we notice that if we make
a definition of a connection similar to that in equation (\ref{connection-1}), 
we will obtain a $U(k+1)$ gauge field, consistent with the interpretation
of a stack of $k+1$ emergent D-branes.  Finally, substituting our
solution into equation (\ref{volume}) we get an answer of the form
$(\text{numerical coefficient})\cdot k + {\cal O}(1/k) ~\text{corrections}$, again
confirming that what we have obtained is a sphere of radius one,
wrapped $k$ (or $k+1$) times.  This wrapping seems to be necessary
to recover full rotational symmetry.

The string theory representation raises the following puzzle: is it possible to make a single
emergent spherical $D4$-brane? In \cite{Castelino:1997rv} this puzzle was phrased
differently: is it possible to separate the $k+1$ branes making up the 
stack and give them different radii?  We take a partial step towards
a positive answer in the next section by giving up local $SO(4)$ invariance.

The generalization to from the four sphere to higher even-dimensional spheres, 
$S^{2k}$ is straightforward. These spheres are constructed  in the same way as the four sphere, $S^{4}$,
by simply using the higher dimensional $\gamma$ matrices (see, for example, 
\cite{Ramgoolam:2001zx} for a review).
$SO(2k)$ symmetry around a point on $S^{2k}$ will be restored in much the same
way that $SO(4)$ symmetry was restored around a point on $S^{4}$, leading
to higher dimensional versions of the noncommutative algebra (\ref{spin-nc}).
Even-dimensional noncommutative spheres have a rich phenomenology (see
for example \cite{Hasebe:2010vp}), which it would be interesting to explore from the 
point of view of our embedding operator.

\section{More examples in $d=5$}
\label{examples}

In this section, we consider two relatively simple co-dimension one hypersurfaces
in $\mathbb{R}^5$, one of which has the topology and the symmetries of $(S^2\times S^2)/\mathbb{Z}_2$,
and the other is a round $S^4$ whose $SO(5)$ symmetry is broken by noncommutativity.

To embed $(S^2\times S^2)/\mathbb{Z}_2$ in $\mathbb{R}^5$, we consider the equation
\be
(1-x_2^2-x_3^2) (1-x_4^2-x_5^2) = x_1^2~.
\label{ex1}
\ee
The noncommutative version of this hypersurface is given by
\ba
X_1&=J^{(1)}_3 \otimes J^{(2)}_3~, \label{s2-times-s2} \\
X_2&=\boldsymbol{1} \otimes J^{(2)}_1~, \nn \\X_3&=\boldsymbol{1} \otimes J^{(2)}_2~, \nn \\
X_4&=J^{(1)}_1 \otimes \boldsymbol{1}~, \nn \\ X_5&=J^{(1)}_2 \otimes \boldsymbol{1}~, \nn 
\end{align}
where the matrices $J_i^{(a)} = L_i^{(a)}/j_a$, while $L_i^{(a)}$
form two irreducible representations of ${su}(2)$:
$[L_i^{(a)}, L_j^{(a)}] =  i \epsilon_{ijk} L_k^{(a)}$,
each with spin $j_a$, $a=1,2$.  It is easy to see that, in the large spin limit,
these matrices satisfy equation (\ref{ex1}).  

At the point $(1,0,0,0,0)$, 
the corresponding embedding operator has two zero eigenvectors,
$V_4\otimes(|j_1\rangle\otimes|j_2\rangle):=V_4\otimes|\alpha_1\rangle$ and  
$\tilde V_4\otimes(|-j_1\rangle\otimes|-j_2\rangle):=V_4\otimes|\alpha_2\rangle$,
where $J_3^{(a)}|m\rangle_a = m|m\rangle_a$.  The local noncommutativity at this point is
\ba
\langle \alpha_1 | -i[X_2,X_3] | \alpha_1 \rangle &=-\langle \alpha_2 | -i[X_2,X_3] | \alpha_2 \rangle = \frac{1}{j_1}~,~ \\
\langle \alpha_1 | -i[X_4,X_5] | \alpha_1 \rangle &=-\langle \alpha_2 | -i[X_4,X_5] | \alpha_2 \rangle = \frac{1}{j_2}
\end{align}
with the expectation values of the other commutators vanishing, and with all
cross-terms between $|\alpha_1\rangle$ and $|\alpha_2\rangle$ vanishing as well for
$j_i > 1/2$.  

The set of matrices (\ref{s2-times-s2})
has the expected $SO(3)\times SO(3)$ symmetry: an action of the
symmetry group on the lower indices of $J^{(a)}_i$ is equivalent
to a conjugation.  However $SO(3)\times SO(3)$ is not a subgroup
of $SO(5)$, so different points on the emergent surface are 
not equivalent and we cannot use symmetry to study zero eigenvectors
of the embedding operator.  Instead, we must resort to numerical analysis.
Preliminary numerical study at various small spins (at most $2$)
shows that the emergent surface gets closer to that in equation (\ref{ex1}) for larger matrices,
and that the embedding operator has two zero eigenvectors everywhere on the emergent surface.
This would imply that the emergent surface locally looks like a direct sum of two noncommutative
spaces described in section \ref{flat}.  It is possible that there
are some points of enhanced symmetry, though we did not find any.
That we get two copies of noncommutative flat space locally is consistent with
$S^2\times S^2$ being a double-cover of the surface  in equation (\ref{ex1}).

A different noncommutative surface is given by
\ba
X_1&=J^{(1)}_3 \otimes J^{(2)}_3~, \\
X_2&=J^{(1)}_3 \otimes J^{(2)}_1~, \nn \\X_3&=J^{(1)}_3 \otimes J^{(2)}_2~, \nn \\
X_4&=J^{(1)}_1 \otimes \boldsymbol{1}~, \nn \\ X_5&=J^{(1)}_2 \otimes \boldsymbol{1}~. \nn 
\end{align}
These five matrices satisfy, in the large spin limit, the equation $\sum_i X_i^2 = 1$.  
Again, at the point $(1,0,0,0,0)$, $V_4\otimes(|j_1\rangle\otimes|j_2\rangle)$ and  $\tilde V_4\otimes(|-j_1\rangle\otimes|-j_2\rangle)$
are zero eigenvectors of the the corresponding embedding operator. 
At this point, the noncommutativity is the same as in the previous example.
Since $SO(5)$ symmetry here is broken to $SO(3)\times SO(2)$, to study the 
whole surface, we resort to numerical analysis, which shows that the embedding operator has two eigenvectors at nearly all
points on the sphere $\sum_i x_i^2 = 1$, except on the circle $x_1=x_2=x_3$,
where the degeneracy is $2j_2+2$.  We can explain the enhanced degeneracy on
the circle as follows:  On this circle, let's take (without loss of
generality) the point (0,0,0,1,0).  The operator $\Lambda_-^2\otimes\boldsymbol{1}+\boldsymbol{1}\otimes L_+^{(2)}$
commutes with $E_5(0,0,0,1,0)$ and generates a basis for its kernel when acting on
$(|\sigma_1,+1\rangle\otimes|\sigma_3,-1\rangle)\otimes(|L^{(1)}_1,+j_1\rangle\otimes|L^{(2)}_3,-j_2\rangle)$
where the notation $|L,l\rangle$ means an eigenvector of operator $L$ with eigenvalue $l$.
Our interpretation is that this corresponds to a stack of two noncommutative spherical
surfaces which `merge' on the circle  $x_1=x_2=x_3$ where, perhaps, the full $SO(4)$ symmetry
is locally restored.  Away from this circle, noncommutativity breaks $SO(4)$ symmetry while
the surface is still a round sphere independent of matrix size.  

These two examples illustrate the rich noncommutative phenomenology that can
be studied using our embedding operators.

\section{Even dimensions}
\label{even}

In this section, we
try to use dimensional reduction of our embedding operator $E_d$
to obtain an embedding operator in even dimensions.
However, we find that this naive attempt does not produce an embedding
operator compatible with the usual construction of the noncommutative 
three sphere $S^3$.  Therefore, we leave even dimensional spaces for future work.

To obtain a guess for the embedding operator in even dimensions,
simply assume that $X_{d}=\boldsymbol{0}$ in equation (\ref{heff}):
\be
E_{d}(x_1,\ldots,x_{d}) 
~=~\sum_{i=1}^{d-2} ~ \left(\sigma^3\otimes\tilde\gamma^i\right)\otimes \left (X_i - x_i \right )
~-~\left(\sigma^1\otimes\boldsymbol{1}\right)\otimes \left (X_{d-1} - x_{d-1} \right )
~-~\left(\sigma^2\otimes\boldsymbol{1}\right)\otimes \left (x_{d} \right )
\ee
It is possible to show that this operator has an eigenvector with eigenvalue
zero only if $x_d=0$ and if another operator, which we would like 
to identify with $E_{d-1}$, has an eigenvector with eigenvalue zero.
This would lead us to propose that
\be
E_{d-1} ~=~\sum_{i=1}^{d-2} ~ \gamma^i \otimes \left (X_i - x_i \right )
~+~i\boldsymbol{1}\otimes \left (X_{d-1} - x_{d-1} \right )~,
\label{heff-even}
\ee
where the $\gamma$ matrices are those for dimension $d-1$,
is a suitable embedding operator in an even dimension $d-1=2n$.
The last term can, equivalently, have a minus sign in front of it.
Notice that the above embedding operator is not hermitian:
This is inconvenient but seems unavoidable.
A potentially interesting observation is that if we take the last dimension, 
$d$, to be time, then the matrix $X_{d}$ would be anti-hermitian and
$E_{d}$ itself would be hermitian.  

The most natural place to test this embedding operator is to take $d-1=4$
and try the matrices corresponding to a noncommutative $S^3$ (see, for example, 
\cite{Guralnik:2000pb,Ramgoolam:2001zx}).
The corresponding embedding operator does not seem to have any
eigenvectors away from the origin $(0,0,0,0)$.  In particular,
for the two smallest presentations of $S^3$, with $N=4$ and $N=12$,
when the corresponding embedding operator is evaluated at a point $(x,y,z,w)$,
its determinant is
$r^6(r^2+8)$ for $N=4$ and $r^{16}(r^2+6)(r^2+4)^3$ for $N=12$.
We have normalized our matrices so that their largest eigenvalue is $1$,
and $r^2=x^2+y^2+z^2+w^2$.  Clearly, the embedding operator has zero
eigenvectors only at the origin $r=0$.  Thus, (\ref{heff-even}) does 
not seem to be the correct operator.

To understand why the embedding operator in equation (\ref{heff-even}) 
does not have the right properties, it is useful to look at equation
(\ref{coherent}).  Let the $X_i$ be a series of representations
of some Lie algebra (such as $su(2)$ for the two-sphere), so scaled that
eigenvalues have a fixed range.  Due to this scaling, the commutators on the right hand side of 
equation (\ref{coherent})  get smaller as the matrices grow.  
This, in turn, guarantees that the width of the coherent state, on the
left hand side of equation (\ref{coherent}),  approaches zero as
the matrices get large.  However, when the embedding operator in equation 
(\ref{heff-even}) is squared, the off-diagonal terms fail to arrange themselves
into commutators, and we cannot make any conclusions about the size of the
coherent state.  
In work \cite{Ishiki:2015saa}, the existence of coherent states
whose width approaches zero as the matrices grow large was used to
define an emergent surface in any number of dimensions at infinite $N$ 
(but not at finite $N$, in contrast to our work).  We suspect that the corrent embedding operator must
lead to an equation similar in structure to (\ref{coherent}),
and this is why (\ref{heff-even}) fails.

%%%%%%%%%%%%%%%%%%%%%%%%%%%%%%%%%%%%%%%%%%%%%%%%%%%%%%%%%%%%%%%%%%%%
%  END MATTER: BIBLIOGRAPHY, ACKNOWLEDGMENTS, ...                  %
%%%%%%%%%%%%%%%%%%%%%%%%%%%%%%%%%%%%%%%%%%%%%%%%%%%%%%%%%%%%%%%%%%%%

\section*{Acknowledgments}

This research was partially
funded by the Natural Sciences and Engineering Research Council of Canada (NSERC).
JLK is grateful to Lior Silberman for helpful discussions.

\bibliographystyle{JHEP}
\bibliography{new}

\providecommand{\href}[2]{#2}\begingroup\raggedright\begin{thebibliography}{10}

\bibitem{Berenstein:2012ts}
D.~Berenstein and E.~Dzienkowski, {\it {Matrix embeddings on flat $R^3$ and the
  geometry of membranes}},  {\em Phys.Rev.} {\bf D86} (2012) 086001,
  [\href{http://arxiv.org/abs/1204.2788}{{\tt arXiv:1204.2788}}].

\bibitem{deBadyn:2015sca}
M.~H. de~Badyn, J.~L. Karczmarek, P.~Sabella-Garnier, and K.~H.-C. Yeh, {\it
  {Emergent geometry of membranes}},
  \href{http://arxiv.org/abs/1506.02035}{{\tt arXiv:1506.02035}}.

\bibitem{Berenstein:2013tya}
D.~Berenstein and E.~Dzienkowski, {\it {Numerical Evidence for Firewalls}},
  \href{http://arxiv.org/abs/1311.1168}{{\tt arXiv:1311.1168}}.

\bibitem{Chatzistavrakidis:2011gs}
A.~Chatzistavrakidis, H.~Steinacker, and G.~Zoupanos, {\it {Intersecting branes
  and a standard model realization in matrix models}},  {\em JHEP} {\bf 1109}
  (2011) 115, [\href{http://arxiv.org/abs/1107.0265}{{\tt arXiv:1107.0265}}].

\bibitem{Polchinski:1998rr}
J.~Polchinski, {\it {String theory. Vol. 2: Superstring theory and beyond}}, .

\bibitem{Steinacker:2010rh}
H.~Steinacker, {\it {Emergent Geometry and Gravity from Matrix Models: an
  Introduction}},  {\em Class.Quant.Grav.} {\bf 27} (2010) 133001,
  [\href{http://arxiv.org/abs/1003.4134}{{\tt arXiv:1003.4134}}].

\bibitem{Steinacker:2011ix}
H.~Steinacker, {\it {Non-commutative geometry and matrix models}},  {\em PoS}
  {\bf QGQGS2011} (2011) 004, [\href{http://arxiv.org/abs/1109.5521}{{\tt
  arXiv:1109.5521}}].

\bibitem{Berezin:1974du}
F.~Berezin, {\it {General Concept of Quantization}},  {\em Commun.Math.Phys.}
  {\bf 40} (1975) 153--174.

\bibitem{Myers:1999ps}
R.~C. Myers, {\it {Dielectric branes}},  {\em JHEP} {\bf 9912} (1999) 022,
  [\href{http://arxiv.org/abs/hep-th/9910053}{{\tt hep-th/9910053}}].

\bibitem{Grosse:1996mz}
H.~Grosse, C.~Klimcik, and P.~Presnajder, {\it {On finite 4-D quantum field
  theory in noncommutative geometry}},  {\em Commun.Math.Phys.} {\bf 180}
  (1996) 429--438, [\href{http://arxiv.org/abs/hep-th/9602115}{{\tt
  hep-th/9602115}}].

\bibitem{Castelino:1997rv}
J.~Castelino, S.~Lee, and W.~Taylor, {\it {Longitudinal five-branes as four
  spheres in matrix theory}},  {\em Nucl.Phys.} {\bf B526} (1998) 334--350,
  [\href{http://arxiv.org/abs/hep-th/9712105}{{\tt hep-th/9712105}}].

\bibitem{Falomir:2009cq}
H.~Falomir, J.~Gamboa, J.~Lopez-Sarrion, F.~Mendez, and P.~Pisani, {\it
  {Magnetic-Dipole Spin Effects in Noncommutative Quantum Mechanics}},  {\em
  Phys.Lett.} {\bf B680} (2009) 384--386,
  [\href{http://arxiv.org/abs/0905.0157}{{\tt arXiv:0905.0157}}].

\bibitem{Ramgoolam:2001zx}
S.~Ramgoolam, {\it {On spherical harmonics for fuzzy spheres in diverse
  dimensions}},  {\em Nucl.Phys.} {\bf B610} (2001) 461--488,
  [\href{http://arxiv.org/abs/hep-th/0105006}{{\tt hep-th/0105006}}].

\bibitem{Hasebe:2010vp}
K.~Hasebe, {\it {Hopf Maps, Lowest Landau Level, and Fuzzy Spheres}},  {\em
  SIGMA} {\bf 6} (2010) 071, [\href{http://arxiv.org/abs/1009.1192}{{\tt
  arXiv:1009.1192}}].

\bibitem{Guralnik:2000pb}
Z.~Guralnik and S.~Ramgoolam, {\it {On the Polarization of unstable D0-branes
  into noncommutative odd spheres}},  {\em JHEP} {\bf 0102} (2001) 032,
  [\href{http://arxiv.org/abs/hep-th/0101001}{{\tt hep-th/0101001}}].

\bibitem{Ishiki:2015saa}
G.~Ishiki, {\it {Matrix Geometry and Coherent States}},
  \href{http://arxiv.org/abs/1503.01230}{{\tt arXiv:1503.01230}}.

\end{thebibliography}\endgroup

\end{document}